\documentclass[sn-nature]{sn-jnl}

\usepackage{graphicx}%
\usepackage{multirow}%
\usepackage{amsmath,amssymb,amsfonts}%
\usepackage{amsthm}%
\usepackage{mathrsfs}%
\usepackage[title]{appendix}%
\usepackage{xcolor}%
\usepackage{textcomp}%
\usepackage{manyfoot}%
\usepackage{booktabs}%
\usepackage{algorithm}%
\usepackage{algorithmicx}%
\usepackage{algpseudocode}%
\usepackage{listings}%
\usepackage{amsmath}

\theoremstyle{thmstyleone}%
\theoremstyle{thmstyletwo}%

\theoremstyle{thmstylethree}%

\usepackage{xcolor} 

\begin{document}

\title[Article Title]{A Method of Measuring TES Complex ETF Response in Frequency-domain Multiplexed Readout by Single Sideband Power Modulation}


\author*[1]{\fnm{Yu} \sur{Zhou}}\email{zhouyu@post.kek.jp}

\author[1,2]{\fnm{Tijmen} \sur{de Haan}}\email{tijmen@post.kek.jp}

\author[1,3,4]{\fnm{Hiroki} \sur{Akamatsu}}\email{hirokia@post.kek.jp}

\author[1]{\fnm{Daisuke} \sur{Kaneko}}\email{dkaneko@post.kek.jp}

\author[1,3,5,6]{\fnm{Masashi} \sur{Hazumi}}\email{masashi.hazumi@kek.jp}

\author[1,2,3]{\fnm{Masaya} \sur{Hasegawa}}\email{masaya.hasegawa@kek.jp}

\author[7]{\fnm{Aritoki} \sur{Suzuki}}\email{asuzuki@lbl.gov}

\author[8]{\fnm{Adrian} \sur{T. Lee}}\email{Adrian.Lee@berkeley.edu}

\affil*[1]{\orgdiv{International Center for Quantum-field Measurement Systems for Studies of the Universe and Particles (QUP)}, \orgname{High Energy Accelerator Research Organization (KEK)}, \orgaddress{\street{1-1 Oho}, \city{Tsukuba}, \postcode{3050801}, \state{Ibaraki}, \country{Japan}}}

\affil[2]{\orgdiv{Institute of Particle and Nuclear Studies (IPNS)}, \orgname{High Energy Accelerator Research Organization (KEK)}, \orgaddress{\street{1-1 Oho}, \city{Tsukuba}, \postcode{3050801}, \state{Ibaraki}, \country{Japan}}}

\affil[3]{ \orgname{The Graduate University for Advanced Studies (SOKENDAI)}, \country{Japan}}

\affil[4]{ \orgname{Netherlands Institute for Space Research (SRON)}, \orgaddress{\street{Niels Bohrweg 4}, \city{Leiden}, \postcode{2333}, \state{CA}, \country{Netherlands}}}

\affil[5]{\orgdiv{Institute of Space and Astronautical Science (ISAS)}, \orgname{Japan Aerospace Exploration Agency (JAXA)}, \orgaddress{\street{3-1-1 Yoshinodai}, \city{Sagamihara}, \postcode{2520222}, \state{Kanagawa}, \country{Japan}}}

\affil[6]{\orgdiv{Kavli Institute for the Physics and Mathematics of the Universe (Kavli IPMU)}, \orgname{University of Tokyo}, \orgaddress{\street{5-1-5 Kashiwanoha}, \city{Kashiwa}, \postcode{2778583}, \state{Chiba}, \country{Japan}}}

\affil[7]{ \orgname{Lawrence Berkeley National Laboratory (LBNL)}, \orgaddress{\street{1 Cyclotron Road}, \city{Berkeley}, \postcode{94720}, \state{CA}, \country{United States of America}}}

\affil[8]{\orgdiv{Physics Department}, \orgname{University of California}, \orgaddress{\city{Berkeley}, \postcode{94720}, \state{CA}, \country{United States of America}}}


\abstract{The digital frequency domain multiplexing (DfMux) technique is widely used for astrophysical instruments 
with large detector arrays. Detailed detector characterization is required for instrument calibration and systematics control. 
We conduct the TES complex electrothermal-feedback (ETF) response measurement with the DfMux readout system as follows. 
By injecting a single sideband signal, we induce modulation in TES power dissipation over a frequency range 
encompassing the detector response. The modulated current signal induced by TES heating effect is measured, 
allowing for the ETF response characterization of the detector. With the injection of 
an upper sideband, the TES readout current shows both an upper and a lower sideband. 
We model the upper and lower sideband complex ETF response and verify the model by fitting to experimental data.
The model not only can fit for certain physical parameters of the detector, such as 
loop gain, temperature sensitivity, current sensitivity, and time constant, 
but also enables us to estimate the systematic effect introduced by the multiplexed readout. 
The method is therefore useful for in-situ detector calibration and for estimating systematic effects during 
astronomical telescope observations, such as those performed by the upcoming LiteBIRD satellite.}



\maketitle

\section{Introduction}
\label{section:1}

The transition-edge sensor (TES) finds diverse applications in fields such as astronomy\cite{Irwin2005,Benford2009,Kelley2009}, 
nuclear and particle detection\cite{Horansky2008a,Horansky2008b,Bacrania2009}, and quantum information
\cite{Rosenberg2007, Miller2003}. Scaling the TES application to large detector arrays is 
highly desired, particularly in the development of astronomical observatories and cosmological experiments.
Large TES arrays require multiplexed readout to achieve acceptable wiring counts and heat load in the system 
\cite{Vaccaro2023,Bender2020,Asavanant2021,Akamatsu2021}.
Digital frequency-domain multiplexed (DfMux) readout is one of the techniques that has been 
extensively employed in ground-based Cosmic Microwave Background (CMB) experiments\cite{Dobbs2009}. 
In CMB experiments which require ultra-low systematic errors \cite{LiteBIRD2023}, characterizing and 
establishing complete model of the TES response \textit{in-situ} with the DfMux readout system is pivotal. 

One widely used technique for TES characterization is the complex impedance measurement\cite{Lindeman2004},
which has been instrumental in measuring TES properties such as temperature sensitivity, current sensitivity\cite{Lindeman2004,Lindeman2012,Zhou2018,Zhou2020}, heat capacity\cite{Martino2014,Takei2008}, internal thermal conductance\cite{Goldie2009,Galeazzi2003,Zhou2018}, 
and diagnosing noise performance\cite{Kinnunen2012,Palosaari2012,Akamatsu2009b}.
When performing such a measurement, an AC perturbation voltage can be added to the DC-bias. 
In general, the resulting current variations are not in phase with the applied voltage perturbations due to the 
finite response time of the TES. The ratio of the perturbation voltage to the measured current variation 
is traditionally interpreted as the TES exhibiting complex impedance. 
However, the TES has no significant reactance. 
Instead, the source of the phase difference is the time-variation in the resistance 
of the TES due to electrothermal feedback (ETF) \cite{Irwin1995}.
In DfMux readout, probing the ETF response of the TES by directly applying an AC voltage bias within the detector 
thermal bandwidth ($<$ hundreds of Hz) is prohibited by the capacitance of the LC filter (with a resonance frequency centered around $\sim$MHz). 
One solution to overcome this problem is to introduce amplitude modulation of the carrier as done 
in Taralli et al \cite{Taralli2019}.
We propose another method of creating power modulation by injecting a single sideband (at frequency $f_c + \delta f$)
in addition to the carrier (at frequency $f_c$), and measuring the complex response in the sidebands.   
In this scenario, the TES ETF effect occurs at the power modulation frequency $\delta f$,
whereas the resulting currents are measured at the upper and lower sideband frequencies $f_c \pm \delta f$. 
In addition to the TES ETF response, we also model the finite impedance of 
the DfMux electric circuit at the sideband frequencies.
The two measured sidebands generated from a single injected sideband 
cannot be described by a single frequency-dependent complex impedance.
Therefore, we adopt the term TES complex ETF response.

The paper is arranged as follows: Section \ref{section:2} provides an overview of the experimental 
setup and device. Section \ref{section:3} introduces the theoretical model of the complex ETF response for TES 
in DfMux readout under sideband power modulation. Measurement data and model fits are presented in 
section \ref{section:4}, with a discussion on possible systematic effects. The conclusion is given
in Section \ref{section:5}.
The application of this method providing a way to monitor the TES gain calibration for CMB observation 
is discussed in a companion paper \cite{dehaan2023}.

\begin{figure}[h]%
\centering
\includegraphics[width=0.9\textwidth]{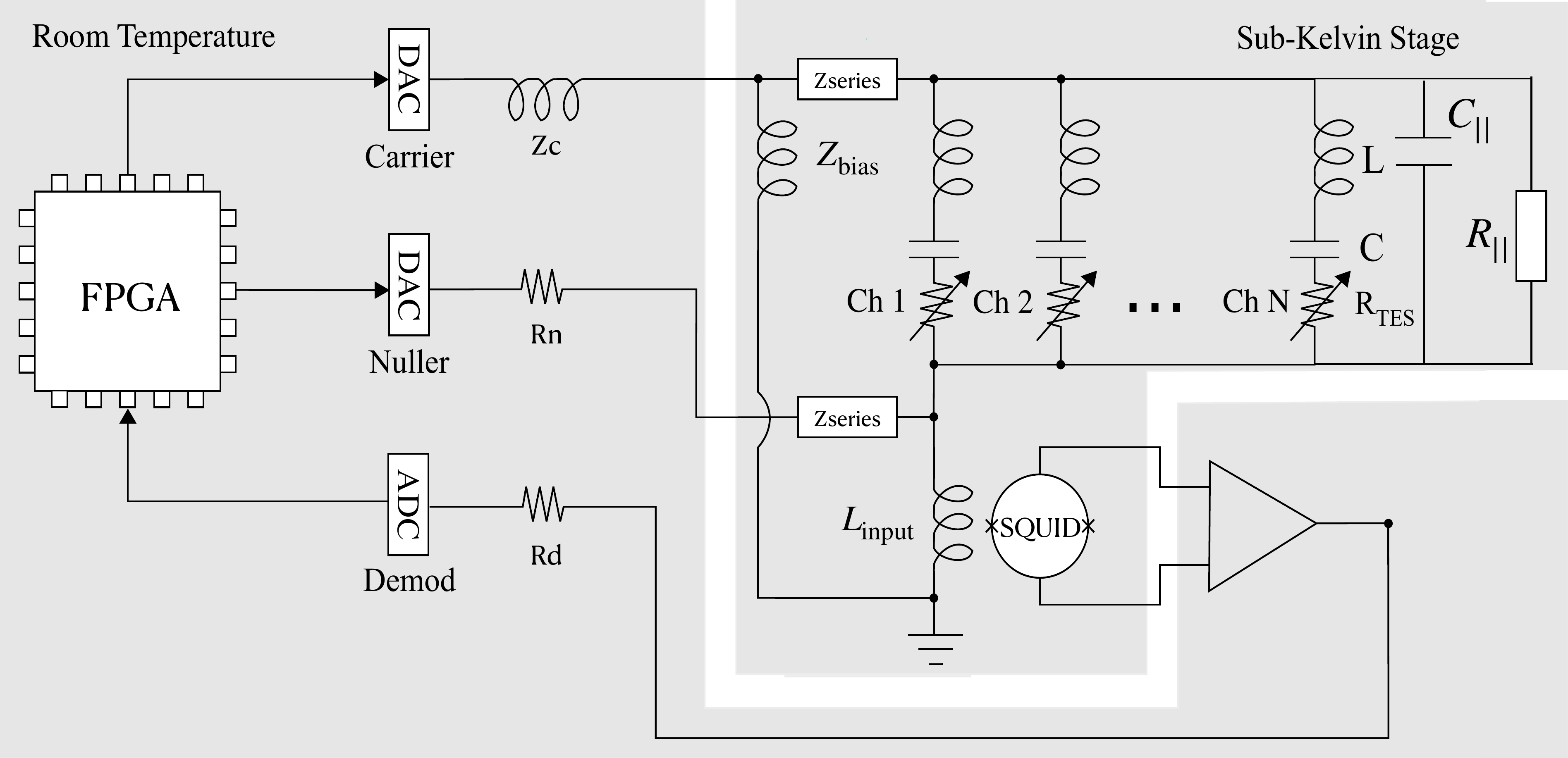}
\caption{A diagram of the digital frequency-domain multiplexed readout circuit. }
\label{fig:1}
\end{figure}

\section{Experimental Setup Overview}
\label{section:2}

The circuit diagram of the TES array and the multiplexed readout is shown in Figure~\ref{fig:1}. 
The cold integrated fMux module (CIMM \cite{Tijmen2020}), including 
the TES array with normal resistance $\sim 0.8 \ \Omega$ and critical temperature $\sim 430\rm\ mK$ (fabricated by commercial superconductor electronics fabrication facility (SeeQC \cite{Suzuki2020})), 
a 40 channel LC resonator (fabricated at LBNL microsystem Laboratory \cite{Rotermund2016}),
a 112-junction series SQUID array amplifier (from STAR Cryoelectronics \cite{Boyd2017}), 
is mounted on a PC board with an A4K $\mu$-metal magnetic 
shield inside an aluminum box, which is installed on the cold stage of a dilution refrigerator. 
The warm electronics is a DfMux version of the ICE readout electronics \cite{Bandura2016}, with a 
field programmable gate array (FPGA) to synthesize the carrier and nuller, and to 
demodulate the signal from the SQUID. Digital Active Nulling (DAN \cite{Tijmen2012}) is used to 
linearize the SQUID response and suppress the input impedance of the SQUID.

\begin{figure}[h]%
\centering
\includegraphics[width=0.9\textwidth]{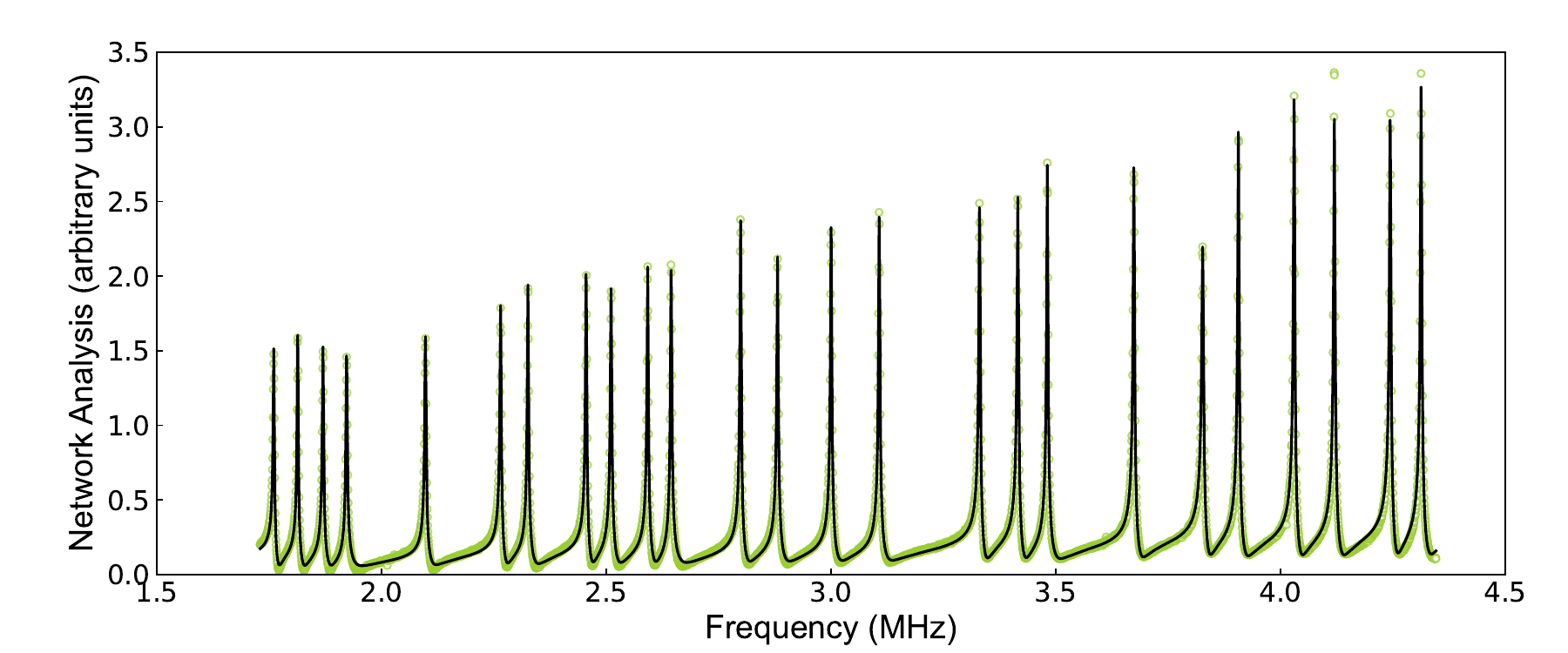}
\caption{Network analysis ($N.A.$) data and the DfMux circuit model fit to the $N.A.$ data.
}
\label{fig:2}
\end{figure}

From the network analysis we find global parameters of the DfMux circuit, i.e. $R_{\rm series} = 0.0834 \rm\ \Omega$,
$L_{\rm series} = 0.0174\rm\ nH $, $L_{\rm bias} = 4.272 \rm\ nH $, $C_{\rm ||} = 3053 \rm\ pF $,
$R_{\rm ||} = 25.1 \rm\ \Omega $, and resonance frequencies for LC combs ($Z_{\rm series} = R_{\rm series} + j\omega L_{\rm series}$, 
$Z_{\rm bias} = j \omega L_{\rm bias}$). The inductance value for LC comb 
is designed to be $L = 59.6\rm\ \mu H$ and fixed as a known parameter in the analysis. Figure \ref{fig:2} 
shows the network analysis data and fit with circuit model.
The DfMux circuit model for network analysis ($N.A.$) is defined as $N.A. =  norm\ |1 + \frac{Z_{\rm }(\omega)}{Z_{bias}}|^{-1}$, 
where
\begin{eqnarray}
Z_{\rm }(\omega, R_{\rm TES_i})|_{i=1,2,...,N} = z_{series} + [ j\omega C_{\rm ||} + \frac{1}{R_{\rm ||}} + \Sigma_{i} (\frac{1}{j\omega L_i + 1/(j\omega C_i) + R_{\rm TES_i}}) ]^{-1}
\end{eqnarray}
is the complex impedance of the DfMux circuit.

\section{Theory on Sideband Complex ETF Response}
\label{section:3}

The TES is held in transition with a voltage bias at carrier frequency $\omega_c$, while an upper sideband 
is supplied at frequency $\omega_c + \delta \omega$. 
In practice, there is not a stiff voltage bias across the TES and rather its voltage bias can be
analyzed as a Th\'evenin equivalent circuit which includes a phase shift caused by series impedance.
However, here we assume perfect voltage bias given by $V  = |V| \cos(\omega_c t)$, and a sideband 
which is a small fraction of the bias voltage $\delta V  = \epsilon |V| \cos((\omega_c + \delta \omega) t)$ 
are supplied to the TES, and apply a complex renormalization to the model transfer function in the end.
We have chosen $\epsilon = 0.02$ as we empirically found it to provide satisfactory signal-to-noise ratio 
while avoiding non-linearity. 
The sideband alternates between constructively and destructively interfering with the carrier, causing power modulation at frequency $\delta \omega$. The resistance responds to this power modulation with some time lag. This causes the demodulated sideband current and voltage to have a frequency-dependent relation in both amplitude and phase. This current-to-voltage transfer function is what we call the complex ETF response.

Assuming a small time-varying perturbation $\delta V $ is produced in TES voltage, and the TES resistance changes by $\delta R $ accordingly.
The TES current $I + \delta I  = (V+ \delta V )/(R+ \delta R )$, can be expanded and reduced to   
\begin{eqnarray}
\delta I  = \frac{\delta V }{R} - \frac{ V \delta R}{R^2}.
\end{eqnarray}

Defining $\alpha = \frac{T}{R} \frac{\delta R}{\delta T} |_{I = I_0}$ as the TES temperature sensitivity, 
and $\beta = \frac{I}{R} \frac{\delta R}{\delta I} |_{T=T_0}$ as the TES current sensitivity, 
the small change in TES resistance can be written as  
$\delta R = \alpha (R/T)\delta T + \beta (R/I)\delta I$.
Assuming that the TES joule power dissipation is in balance with the thermal heat flowing to 
the heat bath, we have the power-balance equation:
\begin{eqnarray}
\frac{ (V+\delta V )^2 }{ R + \delta R } - G|T_0 - T_{\rm bath}| -(G + C \frac{d}{dt} )\delta T = 0.
\end{eqnarray}
Expand equation (3) to first order and substitute $\delta T$, we get
\begin{eqnarray} 
\frac{2 V \delta V }{R} - \frac{V^2 \delta R }{R^2} - \frac{(G + C\frac{d}{dt})T }{\alpha R} (\delta R - \beta \frac{R}{I} \delta I) = 0,
\end{eqnarray}
where $G$ denotes for the thermal conductance between the TES and heat bath, and $C$ denotes for the heat capacity of the TES.

Given that the TES is biased by the carrier at frequency $\omega_c$, the small change in resistance $\delta R$
is only varied by the time-averaging power modulation as a result of the sideband interference.
Combining equation (2) and (4) we can obtain 
\begin{eqnarray}
\delta R = \frac{ <V \delta V>}{<V^2>} \frac{\beta + 2\mathcal{L(\delta \omega)}}{1+\beta + \mathcal{L(\delta \omega)}} R,
\end{eqnarray}
where $\mathcal{L(\delta \omega)} =\mathcal{L}_0/(1+j\delta \omega \tau_0)$, $\mathcal{L}_0 = \alpha P_{\rm e}/(GT)$ is the loop gain of the electrothermal feedback, 
$P_{\rm e}$ is the electrical power dissipation on TES, and $\tau_0 = C/G$ is the intrinsic time constant. 
Take the expression for carrier voltage bias and sideband into equation (5), we find it reduces to
$\delta R (\delta \omega) = \epsilon R \cos(\delta \omega t) \frac{\beta + 2\mathcal{L(\delta \omega)}}{1+\beta + \mathcal{L(\delta \omega)} }$.
Substituting $\delta R (\delta \omega)$ into equation (2), we arrive at the expression 
for the current responses in the upper and lower sidebands, 
$\delta I = \frac{\epsilon |V|}{R} [-\frac{\beta + 2\mathcal{L(\delta \omega)} }{2 (1+\beta+\mathcal{L(\delta \omega)}) } \cos((\omega - \delta \omega )t) + \frac{\beta + 2 }{2 (1+\beta+\mathcal{L(\delta \omega)}) } \cos((\omega + \delta \omega )t) ] $. 
The final form of TES complex ETF response including DfMux circuit transfer function is
\begin{eqnarray}
\mathcal{Y}_{\pm} = \pm \mathcal{Z}_{\rm }^{-1} (\omega_c \pm \delta \omega, R) \frac{\beta + [1 + \mathcal{L(\delta \omega)}] \pm [1 - \mathcal{L(\delta \omega)}] }{2 (1+\beta+\mathcal{L(\delta \omega)}) }, \rm\ 
\end{eqnarray}
$+$/$-$ for\ upper/lower sideband, respectively. $\mathcal{Z} (\omega_c \pm \delta \omega, R)$ 
refers to the complex impedance of the DfMux circuit as defined in equation (1). Note that in this case,
R refers to the TES resistance of the channel of which sideband power modulation is applied, and TES 
resistances of other channels are set as the values they are under their own bias conditions (in our case they are set to zeros). 

An additional phase calibration is necessary for the complex ETF response measurement, due to the fact
that the current implementation of the DfMux system does not provide timestamp or reference associated 
with the zero-point of the phase in the carrier waveform.
We perform this calibration by driving the TES to its normal state. 
A phase rotation $\angle \mathcal{Z}_{\rm }^{-1} (\omega_c + \delta \omega, R_{\rm N})$ 
is then divided from both the upper and lower sideband complex ETF responses. 
Be aware that there is no concept of phase-difference of the lower sideband current response relative to 
the upper sideband voltage bias since the lower and upper sideband are running on different frequencies, 
$\omega_c - \delta \omega$ and $\omega_c + \delta \omega$. 
But since here the lower sideband is triggered simultaneously with the upper sideband, 
and the lower and upper sideband frequency are symmetric relative to the carrier, the initial phase 
of the lower sideband can be identified as the opposite to the reference phase in the upper sideband.
In practice, we find that there is a small residual phase delay that is not accurately captured by
this phase calibration method. However, this residual phase delay is captured by the complex normalization
prefactor in the model. 

\section{Complex ETF Response Measurement and Data Fit}
\label{section:4}

In complex ETF response measurement, we supply a bias voltage $|V_b|$ at carrier frequency $\omega_c$ 
centered at the LC resonance peak and a sideband $\epsilon |V_b|$, sweep 
the frequency $\omega_c + \omega_m$ from $\omega_c + 1\rm\ Hz$ to $\omega_c + 80\rm\ Hz$, 
and measure the demodulated in-phase and quadrature current $\mathcal{I}$ and $\mathcal{Q}$. 
For phase calibration, we first drive the TES to the normal state with a large bias amplitude and fit for an initial phase 
$\psi_{\rm in}$ from the time-ordered $\mathcal{I}$ and $\mathcal{Q}$ data:
\begin{align}
\mathcal{I}(t_i) = |I_n| \cos \psi_{\rm n} + |I_s| \cos ( \omega_m t_i + \psi_{\rm in}) , \\
\mathcal{Q}(t_i) = |I_n| \sin \psi_{\rm n} + |I_s| \sin ( \omega_m t_i + \psi_{\rm in}) .
\end{align}
$I_n$ is the current amplitude measured at the carrier frequency when TES is in normal state, 
$I_s$ is the current amplitude measured for the sideband frequency when TES is normal state,  
and $t_i$ indicates the discrete time stamp.
When TES is in transition, we observe both upper and lower sideband responses in the $\mathcal{I}$ and $\mathcal{Q}$ data
and a simultaneous fit is needed to determine their phase rotations:
\begin{align}
\mathcal{I}(t_i) = |I_t| \cos \psi_t + |I_{+}| \cos ( \omega_m t_i + \psi_{+}) + |I_{-}| \cos ( -\omega_m t_i + \psi_{-}) , \\
\mathcal{Q}(t_i) = |I_t| \sin \psi_t + |I_{+}| \sin ( \omega_m t_i + \psi_{+}) + |I_{-}| \sin ( -\omega_m t_i + \psi_{-}).
\end{align}
$I_t$ is the current amplitude measured at the carrier frequency when TES is in superconducting transition state,
$I_{+}$ and $I_{-}$ are the current amplitudes measured for upper and lower sidebands, respectively, when TES is in superconducting transition state. 
The upper and lower sideband complex response is defined as $\mathcal{Y}_{+} = |I_{+}|/(\epsilon |V_b|) e^{j(\psi_{+} - \psi_{\rm in})}$, 
and $\mathcal{Y}_{-} = |I_{-}|/(\epsilon |V_b|) e^{j(-\psi_{-} - \psi_{\rm in})}$.
Controlling the bath temperature to walk the TES through its superconducting transition, we 
measured the complex responses at $R_{\rm TES}(i)$ for multiple depths in transition. 

\begin{figure}[h]%
\centering
\includegraphics[width=1\textwidth]{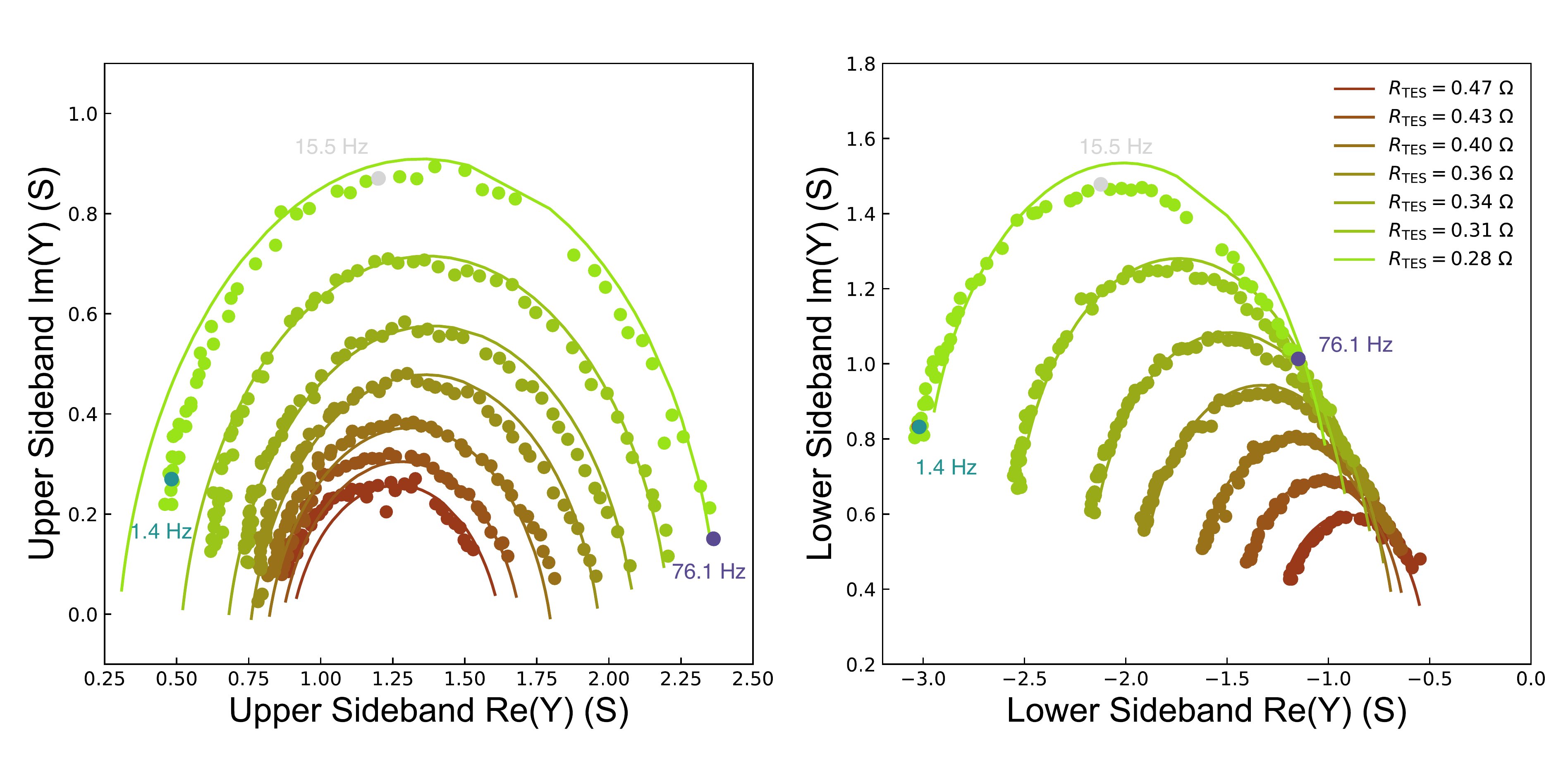}
\caption{Upper sideband ($left$) and lower sideband ($right$) TES complex ETF responses 
measured with multiplexed readout at various depths in transition and the best-fit models.}
\label{fig:3}
\end{figure}

We fit the upper and lower sideband complex ETF responses datasets taken at various $R_{\rm TES}(i)$ 
with model described in section~\ref{section:3} by minimizing the $\chi^2$ function, 
which is the square summation of measured upper and lower sideband TES complex ETF response minus the model divided by average noise level of the data.
Figure \ref{fig:3} shows measured data and model fits for $\mathcal{Y}_{-}$ and $\mathcal{Y}_{+}$ in the complex plane. 

\begin{figure}[h]%
\centering
\includegraphics[width=1\textwidth]{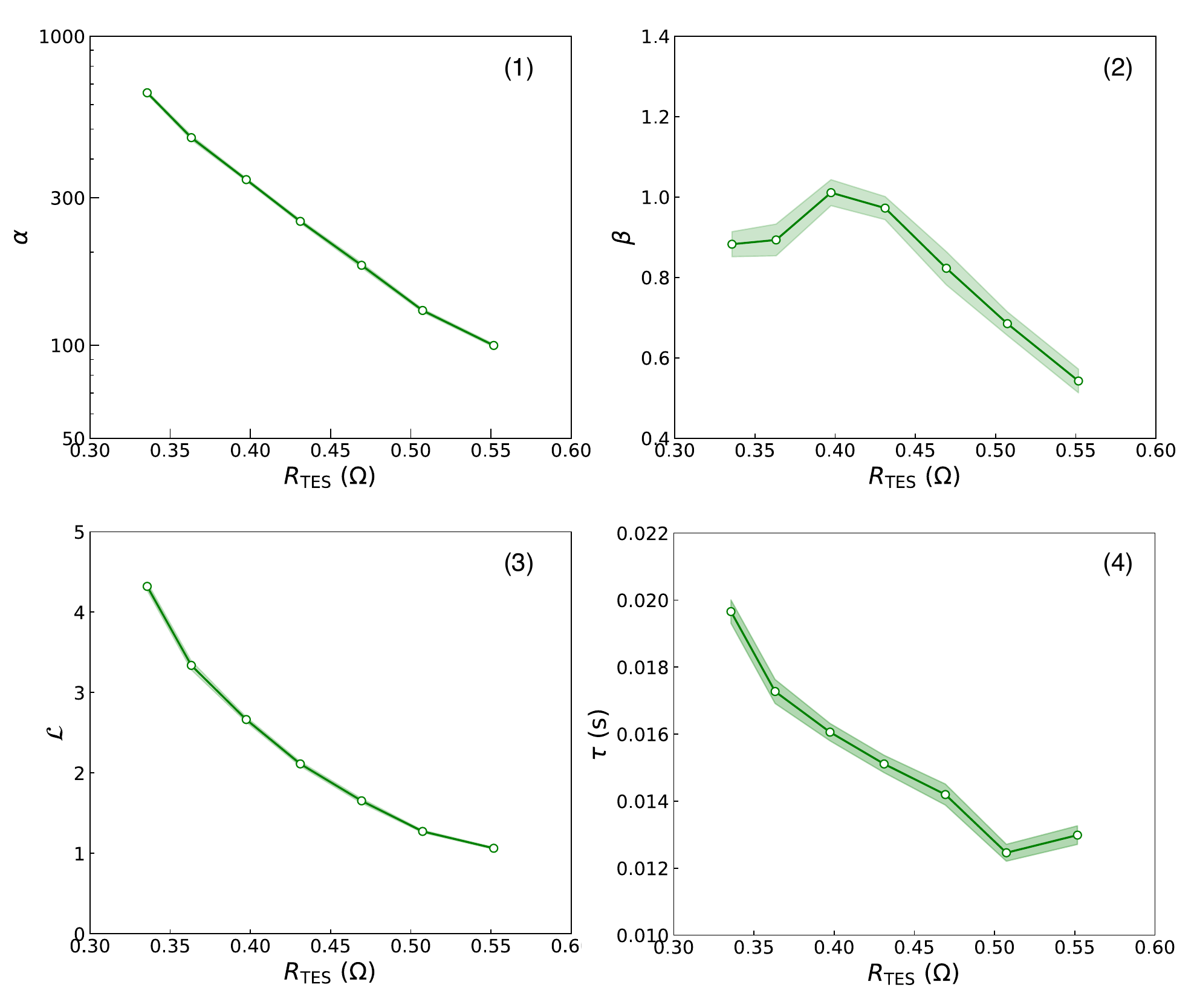}
\caption{Plot of the best-fit TES model parameters versus TES resistances: 
(1) temperature sensitivity; (2) current sensitivity; 
(3) loop gain; (4) time constant. Green dots and shaded regions denote the 
best-fit parameters where $\chi^2_{\rm min}$ are found and statistical errors 
estimated using bootstrapping method, i.e. the standard deviation for the best-fit parameters of the resampled datasets.}
\label{fig:4}
\end{figure}

The $\mathcal{L}_i$, $\beta_i$, and $\tau_i$ are variable for different TES resistances $R_{\rm TES}(i)$, 
while the DfMux circuit parameters, $z_{\rm series}$, $C_{||}$, $R_{||}$, and capacitance of the 
LC filter $C_i$, are global parameters in the fitting procedure. 
In order to determine the statistical uncertainties, we use a bootstrap resampling method; 
the standard deviations are calculated from 100 datasets that were resampled with replacement. The best-fit parameters 
and the uncertainties estimated by bootstrapping are shown in Figure \ref{fig:4}.
When performing simultaneous fitting for the upper and lower sideband complex response data, 
we find a small phase correction is in need applying to $\mathcal{Y}_{+}$ and $\mathcal{Y}_{-}$ with opposite signs, i.e. $\mathcal{Y}_{\pm} e^{\pm j \delta \psi}$.
This uncertainty may be introduced via any common-mode error $\delta \psi$ in determining $\psi_+$ and $\psi_-$ when fitting to the time-ordered data $\mathcal{I}(t_i)$ and $\mathcal{Q}(t_i)$.
We also tried to incorporate a DAN transfer function prefactor of the form $\rm 1/(1 + \omega_m / \omega_0)$.
This did not improve the fit significantly, indicating that this additional model complexity is not necessary
for this work. 

The best-fit equivalent circuit parameters for this TES are $R_{\rm series} = 0.069 \rm\ \Omega$,
$L_{\rm series} = 0.60\rm\ nH $, $C_{\rm ||} = 2753 \rm\ pF $, $R_{\rm ||} = 22.6 \rm\ \Omega $.
Those numbers are not exactly equal to what are derived from network analysis 
as shown at the end of section (2). This is because that the circuit parameters fit with network analysis is 
globally optimized for all channels, and the measurement was done with open-loop setup, where no digital 
active nulling feedback is enabled. On the contrary, during the complex ETF response measurement, 
the digital active nulling feedback is turned on so that the TES is effectively voltage biased, 
and the equivalent circuit parameters we extract from this measurement is dedicated for the specific channel being tested.
Therefore, in terms of modelling the complex response of the TES under constant voltage bias condition, 
the equivalent circuit parameters estimated with this method should be more accurate and reliable than the network analysis.

Figure \ref{fig:4} shows the best-fit 
TES parameters as a function of depth in transition, including the loop gain $\mathcal{L}$, 
time constant $\tau$, current sensitivity $\beta$, and temperature sensitivity $\alpha$, 
which is calculated from $\alpha = \mathcal{L} GT/P_{\rm e}$. 
The thermal conductance $G(T) = G_0(T_0) (\frac{T}{T_{\rm 0}})^{n-1}$ was measured 
independently from the $I-V$ curves by fitting the $P_{\rm TES} (T_{\rm bath}) = \frac{G_0}{n T_0^{n-1}} (T_{\rm TES}^{n} - T_{\rm bath}^{n})$ 
at the same $R_{\rm TES}$ point where current sensitivity is negligible. The best fit results 
are $G_0 = 0.140 \pm 0.004$ nW/K, $T_0 = 428 \pm 1$~mK, and $n = 3.0 \pm 0.2$. With the time constant $\tau = C/G$ measured 
stably in the TES resistance range 0.35$\sim$0.55 $\Omega$ from complex ETF response, we estimate the 
heat capacity of the bolometer is 1.2$\sim$1.6 pJ/K. The measured time constant, thermal conductance, and heat capacity  
are consistent with the designed values \cite{Westbrook2018, Suzuki2013}.
The temperature in TES is calculated according to the thermal conductance function and $T_{\rm bath}$. 
In our companion paper, we have also compared the loop gain, an effective indicator of $\alpha$, derived from the complex ETF response and 
that derived from the $I-V$ measurement, and confirmed that both methods yield consistent results 
at higher bias points where the constant voltage bias condition is still conserved \cite{dehaan2023}. 
The current sensitivity $\beta$ is quite small from this measurement, which is also consistent with the direct derivation of $I-V$ curves 
indicating that $\beta$ is not significantly larger than one or two for this type of TES. 

The fit of simultaneous modeling the upper and lower sideband data is not perfect yet, due to 
the fact that full description of the circuit is very challenging, and further researches are 
still needed to solve the remaining issues in the fitting. Even though, we argue that the 
current result should still be able to provide a less biased measure of detector properties 
compared to single sideband fitting. In fact, we found that our current model can achieve better 
phenomenological fit to either upper sideband data or lower sideband data alone, without fitting to  
the other sideband. Comparing the differences in fit parameters considering either or both sidebands 
can give us an estimation of the systematic bias for this method. We found $\sim$0.1\% difference in the LC filter 
capacitance, $\sim$10\% difference in TES time constant, $\sim$10\% difference in TES loop gain, and $\sim$6\% difference 
in TES current sensitivity. Our method is therefore effective within in this confidence interval, 
and the fit parameters with both sidebands should give us the least systematically biased estimation.
Moreover, the model enables us to further study relevant systematic errors of the detector calibration. 
For example, assuming the carrier frequency
misaligned with the resonance peak frequency by 10 ppm, i.e. 20 Hz deviation from 2 MHz,  
can lead to 1\% error in the loop gain and temperature sensitivity measurement, 4\% error in the current sensitivity 
measurement, and 2\% error in the time constant measurement. Such errors can be detrimental to experiments 
aiming for unprecedented precision, like the proposed LiteBIRD satellite mission.

\section{Conclusion}
\label{section:5}

We propose a novel method for modeling and characterizing the TES properties with the frequency-domain 
multiplexed readout using single sideband power modulation. 
The experiment confirms that TES complex ETF response 
in lower and upper sideband can be determined from the in-phase and quadrature current. 
A theoretical model includes both TES complex ETF response and DfMux circuit transfer function 
is derived under asymmetric sideband power modulation condition. The verification of this model is 
still in a beginning phase. First trials practicing the fit to sideband complex responses suggests 
the method could yield good estimation of the TES loop gain, time constant, temperature sensitivity, 
current sensitivity, and circuit parameters of the multiplexed readout circuit. 
Future efforts such as comparing the independent measurements of detector and circuit properties with 
the parameters fitted from the united model is still needed to justify the proposed method. 
If demonstrated to be effective, the method is potentially very useful in modeling the complex 
response of TESs in frequency-domain multiplexed readout, evaluating cross-talk and current leakage between 
adjacent channels, investigating the impact from any mismatch of carrier and resonance frequency, 
which are essentially valuable for understanding the systematics of experiments 
like the upcoming LiteBIRD CMB satellite project.

\bibliography{TES_DfMux_complexResponsivity}

\end{document}